# Speech Emotion Recognition Using Fine-Tuned DWFormer: A Study on Track 1 of the IERP Challenge 2024

*Honghong Wang, Xupeng Jia, Jing Deng, Rong Zheng*

Beijing Fosafer Information Technology Co., Ltd., Beijing

{wanghonghong,jiaxupeng,dengjing,zhengrong}@fosafer.com

## Abstract

The field of artificial intelligence has a strong interest in the topic of emotion recognition. The majority of extant emotion recognition models are oriented towards enhancing the precision of discrete emotion label prediction. Given the direct relationship between human personality and emotion, as well as the significant inter-individual differences in subjective emotional expression, the IERP Challenge 2024 incorporates personality traits into emotion recognition research. This paper presents the Fosafer's submissions to the Track 1 of the IERP Challenge 2024. This task primarily concerns the recognition of emotions in audio, while also providing text and audio features. In Track 1, we utilized exclusively audio-based features and fine-tuned a pre-trained speech emotion recognition model, DWFormer, through the integration of data augmentation and score fusion strategies, thereby achieving the first place among the participating teams.

## 1. Introduction

Speech emotion recognition (SER) represents a pivotal technology for attaining human-computer interaction [1]. To enhance the efficacy of human-computer interaction, it is imperative to optimize the performance of SER. The conventional approach to SER entails two principal stages: feature extraction and statistical modelling. The former encompasses the extraction of paralinguistic information pertinent to emotional states from the speech signal. This may include prosodic and energy features [2], spectral features [3], and other pertinent characteristics. The latter stage employs a range of general classification models to model and predict these extracted speech features. Common statistical modelling methods [4] include Gaussian Mixture Model (GMM), Hidden Markov Model (HMM), and Support Vector Machine (SVM).

Deep learning-based SER models have significantly improved recognition performance. Leveraging the powerful learning capabilities of neural networks, deep learning not only achieves more complex data classification but also extracts paralinguistic information related to emotions from raw speech signal. By using self-supervised pre-trained speech representation models such as Wav2vec2.0 [5], HuBERT [6], and WavLM [7], emotion-related speech features can be extracted and combined with neural network classifiers to achieve remarkable results.

Corresponding authors: Rong Zheng
This work is supported by the National Key Research and Development Program of China (No.2022YFF0608504)

The current state of SER methods presents a challenge in that they are only capable of predicting discrete emotion labels. This limitation prevents the intensity of the emotion in speech from being accurately reflected [8]. Furthermore, the coexistence of multiple emotions in a single utterance makes it difficult to fully capture the true sentiment by classifying the speech into a single emotion category. In order to address these issues, the IERP Challenge 2024 considers the direct relationship between human personality and emotional expression, incorporating personality traits into emotion recognition.

In Track 1, the organizers employed the use of acoustic and textual features, which were extracted through the pre-trained models such as Wav2vec2.0, to predict scores for eight distinct emotional states. The emotional states included in the study were sadness, happiness, relaxation, surprise, anger, fear, disgust and neutrality. Each emotion score is on a scale of 1 to 5, with 1 representing no emotion and 5 representing the highest level of emotion. The evaluation is based on the root mean squared error (RMSE) of the scores assigned to the eight aforementioned emotions.

A pre-trained SER model, dynamic window transformer (DWFormer) [9], was fine-tuned to predict scores for eight different emotions. Our approach comprises the following stages:

*Model Fine-Tuning*: Building on the pre-trained SER model DWFormer, we froze all parameters except for the last fully connected layer of the classifier, then fine-tuned this fully connected layer using the official dataset.

*Data Augmentation*: To enhance the model's robustness, we augmented the official data by introducing noise from the MUSAN dataset [10]. We established three sets of additive noise probabilities to enhance the robustness of data augmentation strategy.

*Score Fusion*: We combined scores predicted by models trained on various speech features, as well as those obtained from the same feature with different weight parameters, to further enhance model performance. For the above two scores, we used three methods for score fusion: simple averaging, weighted averaging, and maximum value, fusing a total of three scores. In the weighted average method, the weights assigned to the scores of models trained with different features were based on the three smallest RMSE values from the validation set. Similarly, when selecting different training weights for the model, we followed the same RMSE-based selection method.

We fine-tuned the pre-trained SER model DWFormer to adapt with the Track 1 of the IERP Challenge 2024, using WavLM-Large[1], Chinese-HuBERT-Large[2], and Chinese-Wav2vec2-

---

[1]https://huggingface.co/microsoft/wavlm-large

Large[3] as training and testing audio features. The model's performance was significantly improved through noise data augmentation with varying noise addition probabilities and model fusion strategies, leading to a first place ranking in the final submission.

## 2. System Description

DWFormer is a state-of-the-art model in the field of SER. DWFormer is designed to capture the emotional features of speech across different temporal regions and scales. The results on the IEMOCAP [11] and MELD [12] datasets demonstrate that DWFormer outperforms other dominant methods, achieving superior results. Our system is constructed using the DWFormer framework, with all layers except for the final fully connected layer of the classifier being frozen. The classifier is then fine-tuned using the official training data. In contrast to the conventional discrete emotional labeling dataset, the official dataset incorporates a rating for each speech sample across the eight primary emotional categories. The original pre-trained DWFormer is a five-class model designed to recognize happiness, sadness, neutrality, anger, and fear. Its classifier consists of three fully connected layers with rectified linear unit (ReLU) activation functions. In order to accommodate the new emotion classification task, the output of the last fully connected layer was modified to 8 for fine-tuning.

Given the considerable number of parameters in DWFormer and the limited training data provided for the competition, we employed data augmentation on the speech features using the noise subset of the MUSAN dataset to enhance the model's capability. Ultimately, a score fusion module was devised to integrate the outputs of models trained on disparate speech features, in addition to those trained on a single speech feature with varying weight parameters, with the objective of further enhancing the model's performance.

### 2.1. Model Architecture

Figure 1 illustrates the overall structure of the system. Initially, the speech features undergo processing through a data augmentation module for the addition of noise, and are subsequently fed into the Vanilla Transformer [13] Encoder Block. The output comprises the encoded feature representations and the corresponding attention weight matrices. The attention weight matrices are then processed by the Importance Calculation (IC) Module to obtain temporal estimates of the importance of speech emotion information. Subsequently, the importance scores, in conjunction with the encoded output of the original speech features, are conveyed to N stacked DWFormer blocks for the extraction of critical emotional information from the speech. The DWFormer Block consists of the dynamic local window transformer (DLWT) and dynamic global window transformer (DGWT). The DLWT dynamically segments speech features into windows of varying scales, capturing local emotional information within each window, while the DGWT recalculates the significance of each window for global information fusion. The combination of DLWT and DGWT enhances the model's ability to uncover emotion-related paralinguistic information in speech. The output from the DWFormer block is employed by the classifier to predict emotion scores. Ultimately, the model's inferred scores are integrated through the score fusion module.

---

[2]https://huggingface.co/TencentGameMate/chinese-hubert-large
[3]https://huggingface.co/TencentGameMate/chinese-wav2vec2-large

During the training phase, the parameters of the Vanilla Transformer Block, IC Module, DWFormer Block, and Classifier, with the exception of the final layer, are kept unchanged, indicating that the gradients are not updated. The final fully connected layer of the classifier is the sole component subjected to training. The model is trained using a mean squared error (MSE) loss function, as defined by the following equation (1):

$$MSE = \frac{1}{n}\sum_{i=1}^{n}(y_i - \hat{y}_i)^2 \qquad (1)$$

where $n$ denotes the number of samples, $y_i$ represents the true label of the $i$-th sample, and $\hat{y}_i$ denotes the predicted value of the $i$-th sample, respectively.

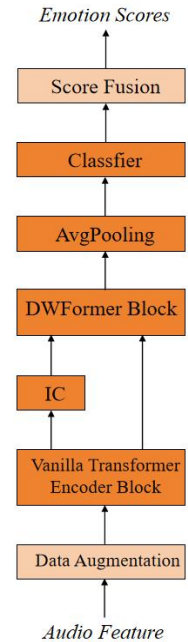

Figure 1: *The flowchat of system architecture. IC represents Importance Calculation module.*

### 2.2. Audio Feature

The audio features utilized in this work, as illustrated in Table 1, encompass three distinct types: WavLM-Large, Chinese-HuBERT-Large and Chinese-Wav2vec2-Large speech representations. These features derived from large pre-trained models, possess enhanced representational capacities, allowing for the capture of finer details in speech, including tone, emotion and semantic information. They are employed across a variety of speech processing tasks, such as speech recognition, speech emotion recognition, and speech synthesis.

Table 1: *Different audio feature. where 'Dim' denotes the dimensionality of frame-level features, 'Params' denotes the number of parameters, 'Hours' denotes the amount of speech data during pretraining.*

| Feature | Dim | Params | Hours |
|---|---|---|---|
| WavLM-Large | 1024 | 300M | 94K |
| Chinese-HuBERT-Large | 1024 | 317M | 60K |
| Chinese-Wav2vec2-Large | 1024 | 317M | 10K |

### 2.3. Data Augmentation

To enhance the model's performance, this study employs data augmentation strategy. Specifically, the noise subset from the MUSAN dataset is used. Given that the audio features encoded by pre-trained models are two-dimensional, the length of the original data is determined by calculating the product of the 0th and 1st dimensions of the feature. Noise samples are then randomly selected from MUSAN dataset. If the length of the noise data exceeds that of the original data, the noise is truncated to match the original length. Conversely, if the noise data is shorter, it is replicated to fit the original length. A noise probability value is established to regulate the random incorporation of noise. If this probability exceeds a randomly initialized value between 0 and 1, noise is added. The specific selection of the noise probability value will be detailed in the experimental section.

### 2.4. Score Fusion

To address discrepancies in performance across models trained with different audio features and the same audio feature under varying training model parameters, we employ score fusion strategy. This approach leverages the strengths of different models to enhance prediction accuracy. Model parameter files are selected based on the RMSE scores obtained from the validation set. We combine the scores using three methods: simple averaging, weighted averaging, and maximum value.

*Simple Averaging*: This method combines scores from models trained with different audio features on the test set, as well as scores from models trained with the same audio feature but different parameters on the test set. Simple averaging can be expressed by equation (2):

$$S_{avg} = \frac{1}{n}\sum_{i=1}^{n} S_i \quad (2)$$

where $n$ represents the number of features or model parameter files. $S_i$ is the score of the model trained using the $i$-th audio feature or the $i$-th model parameter on the test set.

*Weighted Averaging*: This method uses a weighting system based on the RMSE scores of models trained with different audio features and models trained with the same audio feature but different parameters on the validation set to fuse scores on the test set. Weighted averaging is defined by the following equation (3):

$$S_{weighted} = \sum_{i=1}^{n} w_i S_i \quad (3)$$

where $n$ and $S_i$ are consistent with equation (2), $w_i$ is the weight of the model trained using the $i$-th audio feature or the $i$-th model parameter.

*Maximum Value*: In this method, the final score is determined by selecting the highest score from models trained with different audio features and models trained with the same audio feature but different parameters on the test set.

## 3. Experiments and Results

### 3.1. Experimental Setup

We used the official dataset of 2608 audio features for training and validation, and 480 audio features for testing. The dataset was obtained from 163 subjects who watched 16 emotion-inducing videos (including sadness, happiness, relaxation, surprise, fear, disgust, anger, and neutral, with each emotion being randomly presented in two videos). During training, we set the batch size to 4, the initial learning rate to 0.0001, and used the Adam optimizer [14] with a weight decay value of 1e-4. Furthermore, the ReduceLROnPlateau learning rate scheduler [15] was utilized to dynamically adjust the learning rate based on the model's performance during training. In both the training and testing, WavLM-Large, Chinese-HuBERT-Large and Chinese-Wav2vec2-Large representations were employed as audio features. The RMSE of the eight emotion scores serving as the evaluation metric, which is a widely used metric for evaluating the accuracy of a model's predictions. It measures the square root of the average of the squared differences between predicted and actual values. A smaller RMSE value indicates better model performance.

### 3.2. DWFormer Fine-tuning

The pre-trained SER model DWFormer was optimized using various audio features. The results are presented in Table 2. For brief representation, we denote the fine-tuned pre-trained DWFormer as Model_1. Compared to models utilizing WavLM-Large and Chinese-HuBERT-Large features, the model trained with Chinese-Wav2vec2-Large features exhibited the lowest average RMSE on the test set, which indicates that the model using the Chinese-Wav2vec2-Large representation as the audio feature achieves the highest performance.

Table 2: *Test set result of the DWFormer fine-tuning. The arrow signifies that smaller values are preferable.*

| Model | Feature | RMSE ↓ |
|---|---|---|
|  | WavLM-Large | 1.83202 |
| Model_1 | Chinese-HuBERT-Large | 1.90163 |
|  | Chinese-Wav2vec2-Large | **1.79058** |

### 3.3. Data Augmentation

To enhance randomness and diversity in data augmentation, we set random noise addition probabilities of 0.3, 0.5 and 0.8. As illustrated in Table 3, Model_2 represents the fine-tuned DWFormer with noise data augmentation, the experimental results demonstrate that the noise augmentation strategy effectively reduces the RMSE on the test set across three distinct audio features when random noise addition probability is 0.3 and 0.5. Model performance degrades with a random noise addition probability of 0.8. The most favorable outcomes were observed with the Chinese-Wav2vec2-Large feature. Notably, the models trained with all three different audio features achieved the lowest RMSE scores when a random noise addition probability of 0.3 was employed, indicating that a lower random noise addition probability can significantly enhance model performance.

Table 3: *The results on the test set using different random noise addition probabilities.*

| Model | Prob. | Feature | RMSE ↓ |
|---|---|---|---|
| Model_2 | 0.8 | WavLM-Large | 1.98074 |
| | | Chinese-HuBERT-Large | 1.93404 |
| | | Chinese-Wav2vec2-Large | 1.86812 |
| | 0.5 | WavLM-Large | 1.82412 |
| | | Chinese-HuBERT-Large | 1.81857 |
| | | Chinese-Wav2vec2-Large | 1.78335 |
| | 0.3 | WavLM-Large | 1.80782 |
| | | Chinese-HuBERT-Large | 1.77779 |
| | | Chinese-Wav2vec2-Large | **1.77790** |

### 3.4. Score Fusion

Based on the use of noisy data augmentation, a fusion of the scores from models trained with three different audio features and three different model parameters for the same audio feature on the test set was conducted. Three methods mentioned above were employed for score fusion: simple averaging, weighted averaging, and maximum value. In the weighted averaging method, weights were assigned based on the performance of the models with different features on the validation set. The weights were 0.6, 0.2, and 0.2 for the models trained with Chinese-Wav2vec2-Large, Chinese-HuBERT-Large and WavLM-Large, respectively. The same weighting configuration was used for models trained with the same audio feature but different parameters.

Table 4: *The results of score fusion methods using multi-feature models and single-feature models with different parameters on the test set.*

| Model | Fusion Method | RMSE ↓ |
|---|---|---|
| Model_3 | Average | 1.78673 |
| | Weighted Average | 1.80780 |
| | Max | 1.78492 |
| Model_4 | Average | 1.77157 |
| | Weighted Average | 1.77192 |
| | Max | **1.77042** |

As illustrated in Table 4, Model_3 represents the fine-tuned DWFormer with noise data augmentation, utilizing diverse audio features for score fusion. Model_4 represents the fine-tuned DWFormer with noise data augmentation, utilizing the Chinese-Wav2vec2-Large feature and training with distinct model parameters for score fusion. This feature-trained model was selected for score fusion because of its superior performance in previous experiments. 'Average' refers to the simple averaging score fusion method, 'Weighted Average' denotes the weighted averaging score fusion method and 'Max' represents the maximum value score fusion method. Compared to single-feature models, the score fusion methods for different model parameters demonstrated a notable improvement in RMSE on the test set. Multi-feature model score fusion approaches reduce performance Additionally, compared to the simple averaging and the weighted averaging score fusion methods, the maximum value score fusion method yields better results. Limited by the number of submissions, the final ranking result, presented in Table 4, row 5, shows that Model_4 achieved a score of 1.77157 using the average score fusion method, securing the first place among all entries.

## 4. Conclusions and Discussions

This paper presents the system description submitted for Track 1 of the IERP Challenge 2024, which ranked first among all participants. We utilized audio features exclusively, fine-tuning the pre-trained speech emotion recognition model DWFormer, and employed noise data augmentation and score fusion strategies. The experimental results indicate that the model using Chinese-Wav2vec2-Large as the audio feature outperforms those using Chinese-HuBERT-Large and WavLM-Large. Applying noise data augmentation with a randomized noise addition probability of 0.3 further improves the RMSE score compared to probabilities of 0.5 and 0.8. Lastly, among the three different score fusion methods based on two fusion modes, the Maximum Value score fusion method achieved the best performance. Due to submission limits, the final leaderboard result is 1.77157, rather than the system's optimal result of 1.77042.

The aforementioned experiments demonstrated the efficacy and superiority of our system. However, for Track 1, there are still some outstanding issues that require attention. In particular, the text features provided by Track 1 have not been fully exploited, which limits performance. The future of emotion recognition research will likely focus on multimodal emotion recognition, as multimodal approaches have been shown to outperform unimodal ones. The other system, which employs alternative open-source datasets to train an eight-classification SER model and subsequently utilizes official data for fine-tuning, was not pursued due to time constraints. Furthermore, since the official data is labeled with sentiment intensity scores ranging from 1 to 5, the method of using open-source sentiment data to fine-tune the sentiment pre-training model to produce official-like labels and extend the data was not employed. This method could be investigated in future work.